\documentclass[14pt]{article}

\usepackage[dvips]{graphics}
\usepackage{amsmath,amssymb}
\usepackage[usenames]{color}
\def\g{\ensuremath{\mathcal G}}
\def\b{\ensuremath{\mathsf{b}}}
\def\R{\ensuremath{\mathbb{R}}}
\def\C{\ensuremath{\mathbb{C}}}
\def\D{\ensuremath{\mathcal D}}
\def\F{\ensuremath{\mathcal F}}

\def\HF{\ensuremath{\mathcal{H}}}

\def\O{\ensuremath{\mathcal O}}

\def\B{\ensuremath{\mathcal B}}

\def\proof{\textbf{Proof}\\}

\newcommand\partialof[ 2]{\frac{\partial {#1}}{\partial {#2 }}}
\newcommand\coeff [1]{\frac{1}{#1!}}
\newcommand\basisof[1]{\frac{\partial }{\partial #1}}

\newcommand\basispi[2]{\frac{\partial}{\partial \pi^{#1}_{#2}}}
\newcommand\basisq[1]{\frac{\partial}{\partial q^{#1}}}
\newsavebox{\DERIVBOXZLM}
\savebox{\DERIVBOXZLM}[2.5em]{$\sum \hspace{-1em}/$}
 
\def\q{\ensuremath{\hat{q}}}
\def\p{\ensuremath{\hat{\pi}}}
\def\r{\ensuremath{\hat{r}}}

\def\hook{\hbox to 12.5pt{\vbox{\vskip 6pt\hrule width 7pt height .5pt}
         \kern -4pt\vrule height 7pt width .5pt\hfil}}
         
\def\f{\ensuremath{\hat{f}}}

\newcommand{\qed}{$\Box$}

\def\x{\ensuremath{\mathcal X}}

\def\mon{\ensuremath{\hat q^{I_m}_{1_m}\hat\pi_{J_n}\hat r_{1_p}}}

\newtheorem{thma}{Theorem}
\newtheorem{lemma}{\textit{lemma}}

\newtheorem{cor}{Corollary}
\newtheorem{defn}{Definition}

\title{A Full Polynomial n-Symplectic Quantization on $L\R^n$}
\begin{document}
\maketitle 
\begin{center}

{L.K. Norris}\footnote{Department of Mathematics, North Carolina State University, Raleigh, NC 27695-8205,\hfill\break email: lkn@math.ncsu.edu}

\smallskip

{Jonathan D. Brown}\footnote{Department of Mathematics, North Carolina State University, Raleigh, NC 27695-8205,\hfill\break email: 2jdbrown@gmail.com}

\end{center}
\begin{abstract}
n-symplectic geometry,  a generalization of symplectic geometry on the cotangent bundle $T^*M$ of a manifold $M$,  is formulated on the bundle of linear frames $LM$ using the $\R^n$-valued soldering 1-form as the generalized n-symplectic potential.  In this paper we  use n-symplectic geometry on $L\R^n$ to formulate a quantization scheme for a single particle moving in $\R^n$.  By retaining the essence of the standard axioms for quantization on $T^*\R^n$, but adapting them to $L\R^n$, we show it is possible to construct a full polynomial quantization  that is consistent with the Schr\"odinger representation on a $2n$-dimensional subbundle of $L\R^n$.

\end{abstract}

\vfill

Mathematics Subject ClassiÞcations. Primary: 81S99; secondary: 53D99: 

Key words. n-symplectic geometry,  Poisson algebras, quantization, symplectic geometry

\newpage

\section{Introduction}

The "no go" theorems of Groenwold \cite{Gr} and Van Hove \cite{VHove}  show that it is impossible to construct a quantization theory on $T^*\R^n\cong\R^{2n}$, the phase space of a single particle moving in $\R^n$, that is consistent with the Schr\"odinger representation.  In 1946 Groenwold \cite{Gr} proved it is impossible to quantize  all polynomials of $q^i$\  and $p_j$, the canonical coordinates of $\R^{2n}$, in a way consistent with the Schr\"odinger representation.  A few years later in 1951, Van Hove \cite{VHove} proved that it is impossible to quantize $C^{\infty}(\R^{2n})$\ consistent with the Schr\"odinger representation.  This problem has persisted until today and is a subject of study by many authors~\footnote{See the review article on obstructions to quantization by Gotay~\cite{obstructions}.}.

 In this paper we address the question of whether   it is possible to construct a quantization scheme for a single particle moving in $\R^n$ by replacing the underlying symplectic geometry on $T^*\R^n$ with a covering geometry.  Specifically we propose to use n-symplectic geometry on the bundle of linear frames $L\R^n$ in place of symplectic geometry on $T^*\R^n$.  n-symplectic geometry on the bundle of linear frames $LM$ of an n-dimensional manifold $M$ uses the $\R^n$-valued soldering one form $\hat \theta$ as a globally defined n-symplectic potential \cite{LKN1,LKN2,LKN3,LKN4}.  Using the fact that the cotangent bundle $T^*M$ of an n-dimensional manifold $M$ can be considered as the bundle $\R^{n*}\times_{GL(n)} LM$ associated to $LM$, it is not difficult to see that the canonical $\R$-valued symplectic 1-form on $T^*M$ can be defined in terms of the soldering 1-form $\hat\theta$ on $LM$, and that the allowable n-symplectic observables on $LM$ pass to the quotient to define the  polynomial observables of symplectic geometry on $T^*M$~\cite{LKN2}.  In this sense n-symplectic geometry on $LM$ may be considered as a covering geometry for symplectic geometry on $T^*M$.  A brief summary of n-symplectic geometry is presented in section 2.

We take the approach   that Dirac's ``Poisson bracket $\rightarrow$\  commutator" quantization rule is correct,  but symplectic geometry is inadequate to handle a full quantization.    Thus we study Hilbert space based quantizations using n-symplectic geometry.  We focus on the n-symplectic manifold $L\R^n$, the frame bundle of $\R^n$.   The natural observables on any n-symplectic manifold $(LM, d\theta)$ \ are $\otimes^p\R^n$-valued functions that split naturally into two independent classes, namely symmetric Hamiltonian functions $SHF$\ and antisymmetric  Hamiltonian functions   $AHF$.  We  first  show that one may define a full polynomial quantization of the Poisson algebra of symmetric n-symplectic polynomials in $SHF$   that satisfies all the standard quantization axioms except that the basic algebra for the quantization map is not transitive on $L\R^n$.  To remedy this  we reduce the theory to a certain 2n-dimensional subbundle $\B_1$ on which the basic algebra for the quantization map is transitive. We show that the  full n-symplectic polynomial algebra on $L\R^n$ reduces to a reduced n-symplectic polynomial algebra on $\B_1$  that satisfies all the axioms of Hilbert space based quantization.   Hence, there is no obstruction to quantizing $\B_1\subset L\R^n$.

The structure of the paper is as follows.  In Section 2 we present the necessary background materials on n-symplectic geometry, n-symplectic momentum maps and the Hilbert space
we will use for quantization.  In section 3 n-symplectic quantization is defined and in section 4 we derive the basic set needed to  quantize $L\R^n$.  Section 5 contains a direct comparison of the Poisson algebras of polynomials on $T^*\R^n$ and $L\R^n$ in order to better understand  why symplectic quantization fails while n-symplectic quantization succeeds.  Section 6 contains the main results of the paper, namely two explicit polynomial quantizations on $L\R^n$.   Section \ref{conclusions} contains  concluding remarks and suggestions for future work.

It is convienient introduce some notation used throughout this paper.\\
\begin{itemize}
\item $L^2(M)$\ denotes square integrable functions of a manifold $M$.  \\
\item $LM$\ is the bundle of linear frames of a manifold $M$.\\
\item $\otimes_s$\ denotes symmetric tensor product.\\
\item $\otimes_s^p\R^n$\ denotes repeated symmetric tensor product of $\R^n$, $\otimes_s^p\R^n=\R^n\underbrace{\otimes_s\ldots\otimes_s}_{p\ times}\R^n$\\

\item We will use the multi-index notation $f^{I_p}=f^{{i_1}\ldots {i_p}}$.   \\
\item A multi-index on $\p_k$, $\q^i_j$, or $\r_k$ denotes repeated symmetric tensor products over multiple indicies: $\q^{I_a}_{J_b}=\q^{i_1}_{j_1}\otimes_s\ldots\otimes_s\q^{i_a}_{j_b}$.
\item Parenthesis around indicies means symmetrize the indicies.
\item Denote the set of smooth vector fields on a manifold M by $\chi(M)$.
\end{itemize}

\section{Background Material}

\subsection{The Canonical n-Symplectic Manifold LM}
\label{vectorbracket}

In symplectic geometry the canonical symplectic manifold is $P=T^*M$,\ the cotangent bundle of a manifold M.  The symplectic structure, in local symplectic coordinates $(q^i,p_j)$, is given by the differential of the canonical one form $\tilde{\theta}=p_jdq^j$.  To each observable $f\in C^{\infty}(T^*M)$\ one assigns a Hamiltonian vector field by the structure equation

$$ df=X_f \hook d\tilde{\theta}$$
Each symplectic coordinate is $C^{\infty}$ and hence is an allowable observable for $T^*M$.  The corresponding Hamiltonian vector fields are
$$X_{q^i}=-\basisof{p_i},\ \ \ \ X_{p_j}=\basisof{q^j}$$

\begin{defn}
An n-symplectic manifold is a manifold P together with an $\R^n$\ valued non-degenerate two form $\omega=\omega^i\hat{r}_i$.   Here $\{\r_i\}$\ is the standard basis of $\R^n$. 
\end{defn}
An equivalent definition for a polysymplectic manifold is given by Gunther \cite{Gunther}. For n-symplectic geometry the canonical n-symplectic manifold is $P=LM$,\ the linear frame bundle of an n dimensional manifold $M$.  Points in $LM$ are pairs $(m,e_i)$ where $m\in M$ and $e_i$,  $i = 1\dots n$, denotes a linear frame for the tangent space $T_mM$ of the n-dimensional manifold $M$.     Define coordinates on $LM$\ in the standard way. Let $(\tilde{q}^i,U)$\ be a chart on M\ and $\pi:LM\rightarrow M$\ the standard projection to M.  For a point $(m,e_i)\in \pi^{-1}(U)\subset LM$ define coordinates $(q^i, \pi^i_j)$\ by $q^i(m,e_i)=\tilde{q}^i(m)$\ and $\pi^i_j(m,e_k)=e^i(\partial/\partial \tilde{q}^j|_m)$.  The n-symplectic structure is given by the differential of the $\R^n$\ valued soldering one form $\theta=\theta^i\hat{r}_i$\ defined by $\theta(m,e_i)(X)=e^i(d\pi(X))\r_i$ for $X\in  T_{(p,e_i)}LM$.  In local coordinates   $\theta=\theta^i\hat{r}_i=\pi^i_jdq^j\hat{r}_i$. 

 Here the similarity with symplectic geometry starts to differ. The set of observables are $\otimes^p\R^n$\ valued functions on $LM$.  Moreover, the observables are not all of $C^{\infty}(LM)$\ but rather are, locally,  polynomials in the momenta $\pi^i_j$ with coefficients in $C^{\infty}(M)$.  The observables split naturally  into symmetric tensor valued Hamiltonian functions, $SHF$, and totally antisymmetric tensor valued Hamiltonian functions, $AHF$ \cite{LKN1}.  
 For the remainder of this paper we will consider only  $SHF$\ leaving the antisymmetric case to a future work. On LM, all $\otimes^p_s\R^n$\ valued functions, for which there exists a $\otimes_s^{p-1}\R^n$\ valued Hamiltonian vector field, are denoted $SHF^p$.  An element $\hat f\in SHF^p$ has the form  $\hat{f} = \hat{f}^{{i_1}\ldots {i_p}} \hat r_{i_1}\otimes_s \hat r_{i_2}\otimes_s\cdots\otimes_s\hat r_{i_p} $ and will be denoted by the abbreviated notation  $\hat f = \hat{f}^{I_p} \hat r_{I_p}$.    
     Following \cite{LKN1} we assign to each $\hat{f}\in SHF^{p}$\,   an equivalence class $[X_{\f}]$\  of $\otimes_s^{p-1}\R^n$\ valued Hamiltonian vector fields by the structure equation
\begin{equation}
 d\hat{f}^{I_p}=-p!X^{(I_{p-1}}_{\hat{f}} \hook d\theta^{i_p)}
 \label{n-symplectic structure equation}
 \end{equation}
The equivalence class of $\otimes_s^{p-1}\R^n$\ valued vector fields defined by this equation is denoted by $[X_{\f}]=[X^{I_{p-1}}\r_{I_{p-1}}]$.  Although the assignment of Hamiltonian vector fields to observables is unique for rank $p=1$, the symmetrization of the indices in (\ref{n-symplectic structure equation}) introduces a degeneracy that results in the equivalence classes for ranks $p>1$.    In local $(q^i,\pi_k^j)$\ coordinates, the vector fields can be written for $\hat{f}\in SHF^{p}$ as

\begin{equation}
X_{\hat{f}}^{I_{p-1}}=\coeff{p-1}\partialof{\hat{f}^{I_{p-1}b}}{\pi^b_{a}}\basisof{q^{a}}-\left(\coeff{p}\partialof{\hat{f}^{I_{p-1}a}}{q^b}+T^{{I_{p-1}}a}_b\right)\basisof{ \pi^a_b}
\label{preq1}
\end{equation}
where the vertical components   $ T^{{I_{p-1}}a}_b\basisof{ \pi^a_b} $ must satisfy $T^{({I_{p-1}}a)}_b=0  $ but are otherwise arbitrary.
The equivalence classes of Hamiltonian vector fields generated by $SHF$ form a Lie Algebra relative to the bracket defined in \cite{LKN1} as follows.  For $\f\in SHF^p$\ and $\hat{g}\in SHF^q$,\ define the bracket of their corresponding Hamiltonian vector fields by
$$[[X_{\f}],[X_{\hat{g}}]]=[[X_{\f}^{I_{p-1}}\r_{I_{p-1}}],[X_{\hat{g}}^{J_{q-1}}\r_{J_{q-1}}]]\stackrel{def}{=}[X_{\f}^{I_{p-1}},X_{\hat{g}}^{J_{q-1}}]\r_{I_{p-1}}\otimes_s\r_{J_{q-1}}$$
The bracket on the far right hand side is the ordinary Lie bracket of vector fields, and $X_{\f}^{I_{p-1}}$\ and $X_{\hat{g}}^{J_{q-1}}$\ are arbitrary representatives of the equivalence classes $[X_{\f}^{I_{p-1}}]$\ and $[X_{\hat{g}}^{J_{q-1}}]$.  The symmetrization on the upper indices in the bracket destroys the non uniqueness making the bracket independent of choice of representative.  

 These vector fields also preserve the n-symplectic form.
\begin{lemma}
Let $\hat{g}\in SHF^q$\ and $[X_{\hat{g}}^{J_{q-1}}]$\ the corresponding Hamiltonian vector field.  This vector field preserves the n-symplectic form  $d\theta$\ in the sense that,
$$L_{X_{\hat{g}}^{(J_{q-1}}}d\theta^{i)}=0$$
\label{preserve}
\end{lemma}
\proof
The Lie derivative of forms satisfies the familiar relation
$$L_X\omega=X\hook d\omega +d(X\hook\omega)$$
Therefore we have 
$$L_{X_{\hat{g}}^{(J_{q-1}}}d\theta^{i)}=X_{\hat{g}}^{(J_{q-1}}\hook d(d\theta^{i)})+d(X_{\hat{g}}^{(J_{q-1}}\hook d\theta^{i)})=0$$
The last relation reduces to zero since $X_{\hat{g}}^{(J_{q-1}}\hook d\theta^{i)}=-\frac{1}{q!} dg^{J_q}$\ and $d^2=0$.\qquad \qed\\

In contradistinction to the situation on $T^*M$,\ the local coordinates of $LM$\ are no longer observables.  Each observable must be $\otimes^p_s\R^n$\ valued.  The local coordinates define the following basic observables:

\begin{eqnarray}
q^i&\rightarrow& \q^i_j\stackrel{def}{=}q^i\r_j\\
\pi^a_k&\rightarrow& \p_k\stackrel{def}{=}\pi^a_k\r_a
\label{phatqhat}
\end{eqnarray}
The corresponding Hamiltonian vector fields are
$$X_{\q^i_j}=-\basisof{\pi_i^j},\ \ \ \ X_{\p_k}=\basisof{q^k}$$
These are not the only observables that can be construct from the coordinate functions.  In fact one can create  observables from the coordinates, one for each $SHF^p$, $p\geq 1$.  For example,
$q^i\r_j$, $q^i\r_j\otimes_s\r_k$, $q^i\r_j\otimes_s\r_k\otimes_s\r_l$, etc. are all different observables created from the coordinate $q^i$.  

\subsection{The Poisson Bracket on LM}
\label{poissonbracket}
The Poisson bracket on $LM$ plays a fundamental role in our discussion.    We define the Poisson bracket of two symmetric Hamiltonian functions as follows: 
\begin{defn}\label{definition of poisson bracket}
Let $\hat{f}\in SHF^p$\ and $\hat{g}\in SHF^q$.   Then $\{\hat{f},\hat{g}\}\in SHF^{p+q-1}$\ where $\{.,.\}$\  is defined by

$$  \{\hat{f},\hat{g}\} =p! X_{\hat{f}}^{(I_{p-1}}(\hat{g}^{J_q)})\hat{r}_{I_{p-1}}\otimes_s\hat{r}_{J_q} $$
Here $X_{\hat{f}}$\ is any representative of the equivalence class of symmetric Hamiltonian vector fields of $\hat{f}$.    
\end{defn}
In \cite{LKN1} it is shown that this bracket is independent of choice of representative and hence well defined.  The bracket is also anti-symmetric and satisfies the Jacobi identity.  
It is interesting to note that the bracket so defined is the linear frame bundle version of the Schouten-Nijenhuis bracket of symmetric tensor fields on $M$ \cite{LKN3} when restricted to homogeneous elements

The Poisson bracket in n-symplectic geometry is linked to the bracket of Hamiltonian vector fields in a fashion similar to what occurs in symplectic geometry.   The following theorem is an extension \cite{JDB} of the corresponding result in  \cite{LKN1}.  The proof is essentially the same as the proof in  \cite{LKN1}  and will be omitted.

\begin{thma}
Let $\f\in SHF^p$\ and $\hat{g}\in SHF^q$\ then the Hamiltonian vector fields satisfy the relation
$$ CX_{\{\hat{f},\hat{g}\}}=[X_{\f},X_{\hat{g}}] $$
The square bracket is the one defined in section 1 and $C=\frac{(p+q-1)!}{p!q!}$.
\end{thma}

The n-symplectic Poisson bracket and the bracket of equivalence classes of Hamiltonian vector fields are independent of choice of equivalence class.  Hence we will no longer emphasize  the equivalence class and simply refer to a representative $X_{\f}$.  Finally we point out the bracket relation  
\begin{equation}
\{SHF^p,SHF^q\}\subset SHF^{p+q-1}
\label{rankeqn}
\end{equation}
   Both this relation and the arbitrariness in the definition of the n-symplectic Hamiltonian vector fields will play   fundamental roles in the discussion of ideal quantization presented in section \ref{ideal quantization}. 

REMARK:  For the rest of this paper we will use an n-symplectic Poisson bracket that differs by a minus sign from the definition \ref{definition of poisson bracket} above.  We do this in order to allow easy comparison of symplectic and n-symplectic formulas.  In particular, modifying  the definition \ref{definition of poisson bracket} with a minus sign   leads to the formula $\{\q^i_j,\p_k\} = \delta^i_k\r_j$ rather than $\{\q^i_j,\p_k\} = -\delta^i_k\r_j$.

\subsection{Momentum Mappings}
  A generalization of the momentum mapping on $T^*M$  to n-symplectic geometry will be useful for calculating basic sets which will be defined in a later section.  
Following \cite{LKN2} one may define a momentum map for n-symplectic geometry.
\begin{defn}
Let $\Phi:G\times LM\rightarrow LM$\ be an n-symplectic action of a Lie group G on $(LM, d\theta)$.  The mapping $J:LM\rightarrow \g^* \otimes \R^n$\ is a momentum mapping if for each $\xi \in \g$
$$d\hat{J}(\xi)=-\xi_{_{LM}}\hook d\theta $$
where $\xi_{_{LM}}$\ is the infinitesimal generator of the action of G on LM generated by $\xi$, and $\hat{J}(\xi):LM\rightarrow \R^n$\ is defined by 
$$ \hat{J}(\xi)(u)=<J(u), \xi>$$
\end{defn}  
The inner product $<.,.>$\ is the natural extension of the one on $\g\times\g^*$.
$$<\xi,\xi^*\otimes\r_j>=<\xi,\xi^*>\otimes\r_j  $$
Similar to symplectic momentum maps, if $\{\xi_i\}$\ is a basis of \g\ let $\{J_i\}$\ be the $\otimes^p\R^n$\ valued Hamiltonian functions for $(\xi_i)_{LM}$.  Define $\hat{J}$\ by $\hat{J}(\xi_i)=J_i$. This gives a n-symplectic momentum map $J$\ with components $J_i$.

\subsection{Hilbert Spaces}
The quantization we will define is a general Hilbert space based quantization.  In this section we will define the Hilbert spaces needed to discuss the properties of the operators of these quantizations.  Before we define the new Hilbert spaces we mention the standard Hilbert spaces for some quantizations of $T^*\R^n$.  For the metaplectic quantization \cite{obstructions} the Hilbert space is the set of all measurable complex valued square integrable functions of $\R^n$.  The Hilbert space of measurable complex valued square integrable functions of $M$ \ is standardly denoted by $L^2(M, \C)$.  For the Van Hove prequantization \cite{VHove}\ the Hilbert space is essentially $L^2(\R^{2n},\C)$.  The measure in both cases is the one induced by the canonical volume form.

To describe the appropriate Hilbert spaces for $L\R^n$\ we first need to describe an integral for $LM$. The volume for $LM$\ is given by $dV=\Delta(\omega)\wedge (\theta)^n$, where $\omega$ \ is a torsion free connection on LM and $\theta$\ is the soldering one-form. Let $\omega^i_j$\ be the associated one-forms to $\omega$. Define $\Delta(\omega)\stackrel{def}{=}\omega^1_1\wedge \omega^2_1\wedge \cdots \wedge \omega^{n-1}_n\wedge \omega^n_n$\ and $(\theta)^n\stackrel{def}{=}n!\theta^1\wedge \cdots \wedge \theta^n$. This definition of $dV$ is independent of choice of connection\cite{DC}.  For the simple frame bundle $L\R^n$\ a judicious choice of connection gives a more familiar volume $dV=dq^1dq^2\cdots dq^n d\pi^1_1 d\pi^1_2\cdots d\pi^n_n$.  We have suppressed the wedge products in the previous and following equation.  
For the given volume on $L\R^n$\  we make 
the following definitions.  
\begin{defn}
$L^2(L\R^n,\C)$\ is the Hilbert space of measurable square integrable functions from $L\R^n$\ to $\C$.  Let $\phi,\psi \in  L^2(L\R^n,\C)$\ then the inner product is defined by
$$<\phi, \psi >=\int \phi\bar{\psi}dV $$
\end{defn}

\begin{defn}
For each $p\geq 1$ we define a Hilbert space  $\HF^p=\{\psi^{I_p}z_{I_p}| \psi^{I_p}\in L^2(L\R^n,\C)\}$. Here $z_i$\ is the standard basis for $\C^n$. Let $\phi,\psi \in  \HF^p$\ then the inner product is defined by
$$<\phi, \psi >=<\phi^{I_p}z_{I_p}, \psi^{I_p}z_{I_p} >=\sum_{I_p}<\phi^{I_p},\psi^{I_p}>$$
\end{defn}
\begin{defn}
The Hilbert space $\HF$\ is the completion of the direct sum $\tilde{\HF}$ of  $\HF^p$ for all p.
$$\tilde{\HF}=\bigoplus_{p=1}^{\infty} \HF^p$$
The inner product is the standard inner product for a direct sum.  
$$<\oplus _{p=1}^{\infty}\psi^{I_p},\oplus _{q=1}^{\infty}\psi^{I_q}>=\sum_{p=1}^{\infty}<\phi^{I_p}, \psi^{I_p}>$$
\label{innerproduct}
\end{defn}
For each $\psi\in\HF$\ there will be only finitely many non-zero terms, so this inner product is well defined.

\section{Definition of n-Symplectic Quantization}\label{quant}
The definition of n-symplectic quantization we give below is modeled on symplectic quantization. 
We follow the definition of a symplectic quantization given in \cite{obstructions} restricted to the specific symplectic manifold $T^*M$.  A detailed explanation of the motivation behind each condition is given in \cite{obstructions}.  We choose this definition as it is independent of quantization method.

 Let $M$\ be an n dimensional manifold.  Then $(LM,d\theta)$ is an n-symplectic manifold.\ 
\begin{defn}
A prequantization of a Lie subalgebra \O\  of $SHF$\   is a linear map ${\cal Q}$\    where ${\cal Q}$\ takes observables $f \in \O$ \ to symmetric operators on a dense domain D of a Hilbert space \HF\ such that the following hold: \\ 
(1) ${\cal Q}(\{f,g\})=\frac{i}{\hbar}[{\cal Q}(f),{\cal Q}(g)]$\\
(2) If $\r_i\in\O$\  then\ ${\cal Q}(\r_i)=c_i$.  The constants $c_i$ \ are complex numbers.\\
(3) If the Hamiltonian vector field $X_{f}$ of f\ is complete, then ${\cal Q}(f)$\ is essentially self-adjoint on D.
\end{defn}
Here $\{.,.\}$\ denotes the n-symplectic Poisson bracket discussed in section 2.

\begin{defn}
  An n-symplectic basic set of observables  on LM consists of a pair $(B,\b\ )$ where $B$ is a subbundle of LM and  \b\ is a Lie subalgebra of $SHF$\  such that:\\
(4) \b\  is finitely generated,\\
(5) the Hamiltonian vector fields $X_{f}$,\ $f\in\b\ $are complete,\\
(6) {\bf $\b$} is transitive and separating   on B , and \\
(7) \b \ is a minimal Lie algebra satisfying these requirements.
\end{defn}

A set of functions \F  \ on an  n-symplectic manifold $LM$ is transitive  on a subbundle $B$ of LM  if $\{  X_{f}|_B | f\in\F\}$\ span   $B$ .  We say a set of functions separates points on a subbundle $B$ if for $x\neq y\in B$\ there exists an $f\in \F$\ such that $f(x)\neq f(y)$.

\begin{defn}
A quantization on a subbundle $B\subset LM$\ is a prequantization $(\b,{\cal Q})$\ such that for the basic set $(B,\b)$ :\\
(8) ${\cal Q}(\b)$\  acts irreducibly on \HF,\\
(9) ${\cal Q}|_{\b}$\ is faithful, and\\
(10) D contains a dense set of separately analytic vectors for ${\cal Q}(\b)$.\\

Separately analytic has the same meaning here as it does for the symplectic quantization since it is defined in terms of operators.  
\end{defn}
A quantization is said to be a full quantization if $\O=SHF$. 
\begin{defn}
$P(\b)$\ is the polynomial algebra for a basic set $\b$.
\end{defn}
A quantization is said to be a full polynomial quantization if $\O=P(\b)$.\\

\section{Basic Sets for $L\R^n$}\label{basic sets}
Before we compute  basic sets for $L\R^n$\  and certain subbundles of $L\R^n$ we review the basic set for the cotangent bundle of $\R^n$.  In \cite{obstructions} the basic set for $T^*\R^n$\ is the Heisenberg algebra $b_{2n}\subseteq C^{\infty}(\R^n)$

\begin{equation}
\b_{2n}= span\{q^i,p_j,1\}\cong h(2n)
\label{basicset1}
\end{equation}
This basic set consists of the span of the components of the momentum map created by the action of Heisenberg group $H(2n)$\ acting on $\R^{2n}$\cite{STP}.  This is the basic set needed for the Schr\"odinger representation.  

Recall the technique mentioned at the end of section 2.3.  If $\{\xi_i\}$\ is a basis of \g\ let $\{J_i\}$\ be the $\otimes^p\R^n$\ valued Hamiltonian functions for $(\xi_i)_{LM}$.  Define $\hat{J}$\ by $\hat{J}(\xi_i)=J_i$. This gives a n-symplectic momentum map $J$\ with components $J_i$.  This is the procedure we outline below.  Before we begin recall also that $L\R^n\cong \R^n\times Gl(n)$\ and we have a global chart from $L\R^n\rightarrow \R^n\times Gl(n)$\ given by the coordinates $(q^i,\pi^i_j)$.  To compute the n-symplectic momentum map needed to construct the basic set for the frame bundle $L\R^n$\ we define a new group.
\begin{defn}
Define the group 
$$H(L\R^n)=\x(L\R^n)\times^n S^1 $$
with product 
$$(u_1,z^1,\ldots, z^n)\cdot(u_2,w^1,\ldots, w^n)=(u_1+u_2,z^1w^1exp(\frac{i}{2}A^1),\ldots,z^nw^nexp(\frac{i}{2}A^n)) $$
In the above equation $A^k=d\theta^k(u_1,u_2)$\ evaluated at the identity $(0,I)$.
\end{defn}
For every element $(u,\vec{z})$\ we have the inverse $(-u,\vec{z}^{-1})$.  Notice the similarity to the Heisenberg group $H(2n)$.  Given this similarity and following Guillemen and Sternberg \cite{STP} (section 15) we may identify the Lie algebra $h(L\R^n)$\ of this group   with $\x(L\R^n)\times^n \R=\x(L\R^n)\times \R^n$.  The bracket for the Lie algebra is

$$
[(u_1,\vec{v}),(u_2,\vec{w})]=(0,d\theta^1(u_1,u_2),\ldots,d\theta^n(u_1,u_2))=(0,d\theta^i(u_1,u_2)\hat{r}_i)
$$ 
Now we are ready to compute the components of the n-symplectic momentum map for $H(L\R^n)$ \ acting on $L\R^n$.   
Identify $L\R^n$ with a subset of $h(L\R^n)$ \ by $q^i\r_j\rightarrow \q^i_j\rightarrow (X_{\q^i_j},0)$, $\pi^j_k\r_j \rightarrow \p_k\rightarrow (X_{\p_k},0)$.  Also identify the identity $\r_k\rightarrow (0,\r_k)$ and let $H(L\R^n)$\ act on $h(L\R^n)$\ via the adjoint action.  For the elements $\xi=\xi^c_r\hat{q}^r_c+\xi^l\hat{\pi}_l,\ m=m^b_a\hat{q}^a_b+m^k\hat{\pi}_k \in h(L\R^n)$

\begin{equation}
 \xi_{_{L\R^n}}(m)=[\xi,m]=(0,d\theta^i(\xi,m)\hat{r}_i)=d\theta^i(\xi,m)\hat{r}_i=(\xi^b_am^a-m^b_l\xi^l)\hat{r}_b
 \label{bracketh}
 \end{equation}
Given the above identification, a basis for $h(L\R^n)$\ is $\{\q^i_j,\p_k,\r_k\}\rightarrow \{X_{\q^i_j},X_{\p_k}, \r_k\}$.  Using (\ref{bracketh}) we compute the infinitesimal generators of the basis, $(\q^i_j)_{_{L\R^n}}=\basisof{\p^j_i}$,\ $(\p_k)_{_{L\R^n}}=\basisof{\q^k}$,\ and $(\r_k)_{_{L\R^n}}=0$. The components of the momentum map are the Hamiltonian functions for these infinitesimal generators: 
$$ J^a_b=\hat{q}^a_b, J_k=\hat{\pi}_k, J_j=\hat{r_j} $$
We   choose for an   n-symplectic  basic set on $L\R^n$  the pair $(L\R^n,\b_L)$ where $\b_L$ is the span of the components of this n-symplectic momentum map.

\begin{equation}
\b_L=span\{\q^i_j,\p_k, \r_j\} \cong h(L\R^n)
\label{basicset2}
\end{equation}
This is the analogue of the Heisenberg algebra for $L\R^n$. The non-zero Poisson brackets are
$$
\begin{array}{c}
\{\q^i_j,\p_k\}= \delta^i_k\r_j
\end{array}
$$
From the bracket we see $\b_L$\ is a subalgebra of $SHF$.  The Hamiltonian vector fields for the generators of the subalgebra $\b_L$ \ are 
$X_{\hat\pi_k} = \basisq k$,  $X_{\q^j_k} = -\basispi kj$, and $X_{\r_k} = 0$.  
 The integral curves of these vector fields are linear and hence defined for all time.  The set $\b_L$ is finitely generated and since $q^i$\ and $\pi^i_j$\ are global coordinates on $L\R^n$\ they separate points.  Likewise their Hamiltonian vector fields span $T(L\R^n)$.  Thus $\b_L$\ is indeed a basic set for $L\R^n$. The fact that we get the ``hatted" versions of the coordinates instead of the coordinates themselves is a consequence of the fact that all observables on $L\R^n$\ must be $\R^n$\ valued.  

 The fundamental n-symplectic observables defined by the coordinate functions $q^i$ and $\pi^j_k$ are the vector-valued functions $\q^i_j = q^i \r_j$ and $\p_k = \pi_k^l\r_l$.  If we consider the set of all polynomials in $\q^i_j $ and $\p_k $  then the coordinate functions $q^i$ are repeated n-times as the index $j$ on $\q^i_j $ runs from 1 to n. 
  To eliminate this multiplicity  we consider a smaller n-symplectic basic set $(B_1,\b_1)$ where $B_1$ is the subbundle of $L\R^n$ defined as follows.  Let $u^i$, $i=1\dots n$ denote the standard coordinates on $\R^n$ that induce     the corresponding global chart  $(q^i,\pi_k^j)$ on $L\R^n$.  Then $B_1$  is the slice of the coordinate system $(q^i,\pi_k^j)$ obtained by setting
\begin{equation}
\pi^A_j(u) = \delta^A_j\ \ ,\ \ A=2,3,\dots, n\ \ , j = 1,2, \dots, n
\label{definition of B one}
\end{equation}
for all $u\in B_1$.  
It is straight forward to show that $B_1$ consists of points of the form $(p,e_i)$ where $p\in \R^n$ and the frame $(e_i)$ at $p$ is given by
\begin{equation}
e_1 = \alpha\frac {\partial}{\partial u^1}\ \ ,\ \ e_A = \frac {\partial}{\partial u^A}|_p + \mu_A \frac {\partial}{\partial u^1}|_p
\label{points in B one}
\end{equation}
where $\alpha \neq 0$ and $(\mu_A)\in \R^{n-1}$. 

 The coordinates $\pi^i_j$ transform under right translation by $\pi^i_j(u\cdot g) = (g^{-1})^i_k\pi^k_j(u)$.  Using this fact  it is easy to see that the structure group of this bundle is the subgroup  of all $GL(n)$ elements of the form
$
g = 
\left(
\begin{array}{cc}
 a  & \vec b       \\
 \vec 0 &   I  
\end{array}
\right)
$ 
where $a\neq 0$,  $\vec b \in \R^{n-1}$, $\vec 0$ is the (n x 1) zero column vector, and $I$ is the (n-1)x(n-1) identity matrix.  The group multiplication is $(a_1,\vec b_1)\cdot(a_2,\vec b_2) = (a_1a_2,a_1\vec b_2 + \vec b_1)$ and hence $G_1$ is the semi-direct product of the non-zero reals   $ \R^*$ with $\R^{n-1}$. 

For the subalgebra $\b_1$ we choose
\begin{equation}
\b_1 = \hbox{span}_{\R}\{q^i \hat r_1,\hat \pi_k, \hat r_1\}
\label{definition of b sub 1}
\end{equation}
Using the bracket relation $\{\q^i_j,\p_k\}=\delta^i_k\r_j$ it is easy to verify that $\b_1$ is a Lie subalgebra of $\b_L$.  Moreover, by comparing (\ref{basicset1}) and (\ref{definition of b sub 1}) it is clear that $\b_1$ is isomorphic to the Heisenberg algebra.  \bigskip

\noindent   REMARK:  We note that the n-symplectic basic set $(B_1,\b_1)$ is not unique. For example there are n-1 similar subbundles $B_2, B_3,\dots, B_n$  and subalgebras  $\b_2, \b_3,\dots, \b_n$, obtained by choosing the n-1 other possible (n-1) x (n-1) identity submatrices to define the slice of $(q^i, \pi_k^j)$.  Moreover, each of these 2n-dimensional submanifolds is in fact a {\em symplectic manifold}, as can be seen as follows.  Consider for example subbundle$B_1$.  Let $i: \B_1\to L\R^n$ denote the inclusion map.  The coordinates on $B_1$ are $(Q^i, P_k)$
where  $Q^i =q^i\circ i$ and $P_k = \pi^1_k\circ i$.  Then the n-symplectic 2-form $d\hat \theta$ takes the following form on $B_1$:
\begin{equation}
i^*(d\hat\theta) =i^*( d\pi^i_j\wedge dq^j)\hat r_i =i^*( d\pi^1_j\wedge dq^j )\hat r_1 = (dP_j\wedge dQ^j) \hat r_1
\end{equation}
We will show in section \ref{ideal quantization} that these subbundles are also n-symplectic manifolds.


\section{Comparison of the Poisson Algebra of Polynomials for $T^*\R^n$\ vs $L\R^n$}
\label{algebras}
The existence of a ``no-go" theorem for $T^*\R^n$ and the absence thereof for $L\R^n$ stems from the difference in their Poisson algebras, as we will see in a later section.  In this section we explicitly show the differences in the algebras. 

\subsection{The Polynomial Algebra $P(\b_{2n})$}
Recall the basic set $\b_{2n}= span\{q^i,p_j,1\}\subseteq C^{\infty}(\R^n)$\ for $T^*\R^n$\ from equation (\ref{basicset1}).  Then $P(\b_{2n})$\ is just the set of polynomials of the variables $(q^i,p_j)$, and the bracket is the standard Poisson bracket for $T^*\R^n$.

Computing a few important Poisson brackets we see
\begin{eqnarray}
\{q^i,p_j\}&=&\delta^i_j \nonumber \\
\{(q^i)^a,(q^j)^b\}&=&0  \nonumber \\
\{(p_i)^a,(p_j)^b\}&=&0  \nonumber \\
\{(q^i)^2,(p_j)^2\}&=&4\delta^i_jq^ip_j  \nonumber \\
\{(q^i)^3,(p_j)^3\}&=& 9\delta^i_j (q^i)^2(p_j)^2  \nonumber \\
\{(q^i)^2p_b,q^a(p_j)^2\}&=&3\delta_b^a(q^i)^2(p_j)^2
\end{eqnarray}
The notations $(q^i)^a$\ and $(p_j)^a$\ denote  $a$-fold\ products in the underlying commutative algebra $P(\b_{2n})$.  The last two relations are the Poisson relations that lead to the Groenwold obstruction.  Notable subalgebras of $P(\b_{2n})$\ are polynomials of degree two or less and the affine subalgebra, which is the subalgebra of all polynomials linear in $p_j$.  The Poisson algebra $P(\b_{2n})$ has no non-trivial ideals and satisfies 
$$[P(\b_{2n}),P(\b_{2n})]=P(\b_{2n}) $$

\subsection{The Polynomial Algebra $P(\b_1)$}
Recall from equation (\ref{definition of b sub 1})\ that  $\b_1=span\{\q^i_1,\p_k,\r_1\}$\  and consider the Poisson algebra $P(\b_1)$ with bracket defined in section \ref{poissonbracket}.  Elements of $P(\b_1)$\ are polynomials of $(\hat{q}^i_1, \hat{\pi}_k, \hat{r}_1)$ and hence are $\otimes_s^m \R^n$\ valued functions on $L\R^n$, $m$\ being the degree of the polynomial.  A typical monomial looks like $ \q^{I_n}_{1_n}\p_{K_m}\r_{1_l} \in SHF^{n+m+l}$.  
Here the multiplication is the symmetric tensor product.  For example,  $\hat{q}^i_1\hat{\pi}_k\equiv\hat{q}^i_1\otimes_s\hat{\pi}_k$, and $\hat r_{1_n} = \overbrace{\hat r_1\otimes_s\dots\otimes_s\hat r_1}^{n\ \  times}$. 

Using the n-symplectic Poisson bracket defined in section \ref{poissonbracket} we compute (see the appendix) some relevant brackets.
\begin{eqnarray}
\{\q^i_1,\p_j\}&=&\delta^i_j\r_1 \nonumber \\
\{\q^i,\p_j\p_k\} &=&  \delta^i_j \p_k\r_1 + \delta^i_k\p_j\r_1 \nonumber \\
\{\q^i\q^j,\p_k\} &=&  (q^i\delta^j_k + q^j\delta^i_k)\r_1\r_1  \nonumber \\
\{(\q^i_1)^a,(\q^j_l)^b\}&=&0  \nonumber \\
\{(\p_i)^a,(\p_j)^b\}&=&0  \nonumber \\
\{(\q^i_1)^2,(\p_j)^2\}&=&4\delta^i_j\q^i_1\p_j\r_1  \nonumber \\
\{(\q^i_1)^3,(\p_j)^3\}&=& 9\delta^i_j (\q^i_1)^2(\p_j)^2\r_1 \nonumber \\
\{(\q^i_1)^2\p_b,\q^a_1(\p_j)^2\}&=&3\delta_b^a(\q^i_1)^2(\p_j)^2\r_1 
\label{computedbrackets}
\end{eqnarray}

Notice the similarity to the symplectic Poisson brackets above. The main difference is the $\otimes_s^p\R^n$ \ valued rank.  We also have for any two polynomials $\hat{f},\hat{g}\in P(\b_L)$\ 
$$\{\hat{f}\otimes_s^{I_k}\r_{1_k},\hat{g}\otimes_s^{J_l}\r_{1_l}\}=\{\hat{f},\hat{g}\}\otimes_s^{M_{k+l}}\r_{1_{k+l}}$$
The multi-index $M_{k+l}$\ is $(I_k,J_l)$.

\begin{defn}  For each $\lambda\geq 1$ we denote by $SPF^{\lambda}$ the span over $\R$ of monomials of the form $\q^{I_n}_{1_n}\p_{K_m}\r_{1_l} \in SHF^{\lambda}$ where $\lambda = {n+m+l}$.   
\end{defn}

$SPF^{\lambda}$, the subspace of symmetric homogeneous polynomial functions in $\q^i_1$, $\p_j$ and $\r_1$ of n-symplectic rank $\lambda$, is a subset of  $SHF^{\lambda}$, the symmetrical Hamiltonian functions of rank $\lambda$.   Let $P^k = \oplus_{j=k}^\infty SPF^j$.  Then it is clear that $P(\b_1) = \oplus_{k=1}^\infty SPF^k$ and  $SPF^1 = \b_1$.  Hence
\begin{equation}
P(\b_1) = \b_1 \oplus  P^2 \qquad                       \hbox{where}\quad P^2 = \oplus_{k=2}^\infty SPF^k
\end{equation}
Moreover, because of equation (\ref{rankeqn}) we have
\begin{equation}
\{SPF^k,SPF^l\} \subset SPF^{k+l-1}
\label{the rank condition on brackets}
\end{equation}
As a result $P^2$ is a Lie ideal in the n-symplectic algebra $P(\b_1)$.  In fact   (\ref{the rank condition on brackets}) implies that each $P^k$ is a Lie ideal for   $k\geq 2$.

\section{Go theorem for $L(\R^n)$} \label{ideal quantization}\label{go}

\begin{thma}
There exists a full polynomial quantization of the polynomial algebra $P(\b_1)$\ on the subbundle $\B_1$ of $L\R^n$. 
\end{thma}
We prove this existence theorem by giving two examples.  We work initially  on $L\R^n$, and then show how the results descend to the subbundle $\B_1$ to define a proper quantization. 

In the previous section we showed that 
$$ P(\b_1)=\b_1+ P^2$$
where $+$\ represents semi-direct sum with bracket given by 
$$\{(\xi_1,\eta_1),(\xi_2,\eta_2)\}= (\{\xi_1,\xi_2\},\{\xi_1,\eta_2\}-\{\xi_2,\eta_1\}+\{\eta_1,\eta_2\}\}$$
 Thus we can obtain a full quantization of $P(\b_1)$ by quantizing  $\b_1$\  and setting ${\cal Q}(P^2)=0$. This is the approach taken by Gotay \cite{obstructions} when he exhibited a quantization of $T^*\R_+$.  We quantize $\b_1$ using the following standard Schr\"odinger quantization:
\begin{eqnarray}
{\cal Q}(\q^i_1)&=& q^i\\
{\cal Q}(\p_k)&=&-i\hbar\basisof{q^k}\\
{\cal Q}(\r_1)&=&1
\end{eqnarray}
 It is clear that this map satisfies the definition of a prequantization.  The Hilbert space is $L^2(\R^n,\C)$\ and the domain is the Schwartz space of all $C^{\infty}$\ rapidly decreasing functions.  The quantization of $\b_1$\  is  faithful and is   the  Schr\"odinger representation.  Irreducibility of ${\cal Q}(\b_1)$\ follows from the fact that the Schr\"odinger representation $\{q^i,\basisof{q^k}\}$ \ is irreducible on $L^2(\R^n,\C)$.  Likewise the operators ${\cal Q}|_{\b_1}$\ are essentially self adjoint on \D.   Moreover it is known that the Hermite polynomials form a dense set in \D\  of separately analytic vectors for the Schr\"odinger representation.

Finally we must address the fact that the subalgebra $\b_1$ is transitive on the subbundle $\B_1$ and does not span all of $L\R^n$.  We need   to reduce the quantization scheme to a quantization scheme on the subbundle $\B_1$.  Let $i:B_1\to L\R^n$ denote the inclusion mapping.
\begin{thma}
The pull-back of the n-symplectic polynomial algebra $P(\b_1)$\ to the subbundle $\B_1$  under the inclusion mapping     is an n-symplectic algebra with respect to the pull-back $i^*(d\hat \theta)$ of the canonical n-symplectic form $d\hat \theta$ on $L\R^n$. 
\end{thma}
\noindent \proof  Let   
 $\tilde P(\b_1) = i^*(P(\b_1))$.  Assume that the restrictions of all Hamiltonian vector fields $X_{\hat f}^I$ to points of $\B_1$ for all elements $\f\in  P(\b_1)$ are tangent to $\B_1$.  Thus for $\hat f\in P(\b_1)$ and for $u\in \B_1$ we write $X_{\hat f}^I(i(u)) = i_*(\tilde X_{\hat f}^I(u))$ where  $\tilde X_{\hat f}^I$  are  vector fields on $\B_1$.  Using the notation $\tilde f^{Ij}=i^*(\hat f^{Ij})$ and  $d\tilde\theta^j = i^*d\theta^{j}$  it is then straightforward to show that
\begin{equation}
d\tilde f^{Ij} = -p \tilde X_{\hat f}^{(I}\hook d\tilde \theta^{j)}
\end{equation}
where $p$ is the degree of the polynomial $\f$.    The theorem now follows from the following lemma.
\bigskip
\begin{lemma} \label{TheLemma}
The Hamiltonian vector fields $X_{\hat f}^I$ for all elements $\f\in \tilde P(\b_1)$ can be chosen so that their restrictions to $\B_1$ are tangent to $\B_1$.
\end{lemma}
\proof
Coordinates on the subbundle $\B_1$ are $(Q^i,P_j)=(q^i\circ i, \p^1_j\circ i)$.  Hence it is clear that the restrictions  to $\B_1$ of the  Hamiltonian vector fields $X_{\q^i_1} = \basispi 1i$, $X_{\p_j} = \basisq j$ and  $X_{\hat r_1} = 0$  are tangent to $\B_1$, and therefore the restrictions  to $\B_1$ of all Hamiltonian vector fields for all elements of $\b_1$    are tangent to $\B_1$.

Consider next an arbitrary monomial in $P^2$ of the form $\hat f = \mon$, where $m+n+p\geq 2$.   A monomial of this form is an allowable n-symplectic observable (see Section 2) which we express, using $\hat q^i_1 = q^i\hat r_1$ and $\p_k = \pi^j_k\hat r_j$,  in the form $\hat f =   f^{P_\alpha}\hat r_\alpha$ where $\alpha= m+n + p$.  Then
\begin{equation}
  f^{P_\alpha}=  f^{L_mM_pN_n} = q^{I_m}\delta^{(L_m}_{1_m}\delta^{M_p}_{1_p}\pi^{N_n)}_{J_n}
\label{monial n-sym form}
\end{equation}
The n-symplectic Hamiltonian vector fields for such a monomial have the form given in equation (\ref{preq1}).   Explicitly we have\begin{equation}
X_{\hat{f}}^{P_{\alpha-1}}=\coeff{(\alpha-1)}\partialof{{f}^{P_{\alpha-1}b}}{\pi^b_{a}}\basisof{q^{a}}-\left(\coeff{\alpha}\partialof{{f}^{P_{\alpha-1}a}}{q^b} \right)\basisof{ \pi^a_b}
\label{monomial HVF}
\end{equation}
where we have the freedom to add   vertical components of the form $T^{{P_{\alpha-1}}a}_b\basisof{ \pi^a_b}$ where the   $T^{{P_{\alpha-1}}a}_b  $ must satisfy $T^{{(P_{\alpha-1}}a)}_b=0  $ but are otherwise arbitrary.  Moreover the addition of  such   terms does not affect   Poisson brackets.
It is clear that the non-vertical components involving $\basisof{q^a}$ restrict to vector fields tangent to $\B_1$, so we need only consider the vertical components.  Using the form for $ f^{P_{\alpha -1}a}$ from equation (\ref{monial n-sym form}) in the vertical components, and expanding out the symmetrization  on the indices $L_m M_p N_n$ in $\delta^{(L_m}_{1_m}\delta^{M_p}_{1_p}\pi^{N_n)}_{J_n} $, one finds that the summing index "$a$" in the vertical components in (\ref{monomial HVF}) will fall on a term of the form $\delta^a_1$ or $\pi^a_j$.  In the case that the summing index "$a$" falls on a term $\delta^a_1$, then carrying out the sum on $\delta^a_1\basispi ab$ results in a basis vector of the form $\basispi 1b$, which also restricts to a vector field tangent to $\B_1$.  Finally in the event that the summing index 
"$a$"   falls on a term of the form   $\pi^a_j$, then the corresponding vertical component will have the form
\begin{equation}
\left(  \frac {\partial q^{I_m}}{\partial q^b}\delta^{K_{l-1}}_{1_{l-1}}\delta^{L_m}_{1_m}  \pi_{J_{n-1}}^{N_{n-1}}\delta^k_1  \pi^a_j   \right)\basisof{ \pi^a_b}
\label{trouble term}
\end{equation}
For each term of this kind we add a term of the form
\begin{equation}
T^{{P_{\alpha-1}}a}_b\basisof{ \pi^a_b} =-2 \left(  \frac {\partial q^{I_m}}{\partial q^b}\delta^{K_{l-1}}_{1_{l-1}}\delta^{L_m}_{1_m}  \pi_{J_{n-1}}^{N_{n-1}}\delta_1^{[k}  \pi^{a]}_j   \right)\basisof{ \pi^a_b}
\end{equation}
which is explicitly anti-symmetric in the indices $k$ and $a$, which implies $T_b^{{(P_{\alpha-1}}a)}=0  $.  The net result is to interchange the indices $k$ and $a$ in equation (\ref{trouble term}), which in turn results in another basis vector of the form $\basispi 1b$ when the sum on index "$a$" is carried out.  Hence all terms in the Hamiltonian vector fields for the arbitrarily chosen monomial in $P^2$ can be chosen so that their restrictions to $\B_1$ are tangent to $\B_1$.  \qed

 On the n-symplectic manifold $(\B_1,d\tilde\theta^i\r_i)$   we consider the basic algebra 
\begin{equation}
\tilde \b_1 = span_{\R}(\hat Q^i_1,\hat \Pi_j,\r_1)
\label{b1 on B1}
\end{equation}
where $\hat Q^i_1 = \q^i_1\circ i$ and $\hat \Pi_j = \p_j\circ i$.  Denoting by $\tilde P(\tilde \b_1)$ the n-symplectic algebra of all polynomials formed from $\tilde \b_1$, we have the decomposition
\begin{equation}
\tilde P(\tilde \b_1) = \tilde \b_1 + \tilde P^2
\label{polynomial algebra on B one}
\end{equation}
Quantize this polynomial algebra as we did above for $P(\b_1)$ by defining ${\cal Q}(\tilde P^2) = 0 $ and
\begin{eqnarray}
{\cal Q}(\hat Q^i_1)&=& Q^i\\
{\cal Q}(\hat \Pi_k)&=&-i\hbar\basisof{Q^k}\\
{\cal Q}(\r_1)&=&1
\label{quanitzation of tilde b one on B one}
\end{eqnarray}

 \begin{cor}
There is no Groenwold van Hove type obstruction for quantizing the 2n-dimensional subbundle $\B_1$ of $L\R^n$.
\end{cor}  
By Groenwold van Hove type obstruction we mean an obstruction to quantization that:
\begin{itemize}
\item arises as a consequence of the irreducibility condition and the Poission bracket goes to commutator condition and \\
\item requires a restriction of the quantization to a subalgebra of observables to correct. \\
\end{itemize}

\subsection{Another Full Polynomial Quantization}
Let $A^i$, $i = 1\dots n$ denote $n$ real numbers. Another full quantization is given by the map ${\cal Q}(P^3)=0$\ and 

\begin{equation}
\begin{array}{lcl}
{\cal Q}(\r_1)=1& &{\cal Q}(\q^{i}_{1}\p_k)=A^{i}\pi_k^1  \\
{\cal Q}(\q^i_1)= q^i & & {\cal Q}(  \p_j\p_k      ) =  \pi^1_j\pi^1_k \\
{\cal Q}(\p_k)=-i\hbar \basisof{q^k} && {\cal Q}(\p_{k}\r_1)=0 \\
{\cal Q}(\q^{i}_{1}\q^j_1)=A^iA^j& &{\cal Q}(\r_1\r_1)= 0\\
{\cal Q}(\q^{i}_{1}\r_1)=0 & &\\
\end{array}
\end{equation}
  This quantization is tedious but easy to check.
Notice that when restricted to the basic set $\b_1$ this quantization is the same as the previous one.  Thus the map ${\cal Q}$ satisfies the definition of a quantization given in section \ref{quant}.


\section{Conclusion}\label{conclusions}

To avoid the obstructions to quantizing the canonical symplectic manifold $T^*\R^n$\ something must change:  either Dirac's ``Poisson bracket $\rightarrow$\  commutator" quantization rule or the underlying setting of quantization, or both.  In this paper we have shown that if one retains the Dirac quantization rule but replaces symplectic geometry on $T^*\R^n$ with n-symplectic geometry on $L\R^n$, then one can define a full polynomial quantization.  
The motivation for this change is that the symplectic geometry of polynomial observables on $T^*\R^n$\ is induced from the n-symplectic geometry of $L\R^n$.  Hence, n-symplectic geometry is a natural choice of a larger geometry in which to base quantization.  The n-symplectic geometry of $L\R^n$\ allows for the existence of ideals in the Poisson algebra of polynomials, and consequently allows $L\R^n$\ to support full polynomial quantizations.  These ideals are, however,  absent in the Poisson algebra of polynomial observables on $T^*M$.  The authors believe that the existence of ideals in the Poisson algebra is a necessary and sufficient condition in order to be able to define a full quantization.  With that conjecture in mind it is important to notice that ideals exist in the Poisson algebra of polynomial observables of $LM$\ for all M with dim(M)$ > 1$. 

It is known \cite{GeomS}   that the bundle of linear frames $LM$ of an n-dimensional manifold $M$ is an open dense submanifold of the  Whitney sum of n copies of the tangent bundle.   The subbundles  $\B_i$ discussed at the end of section 4 can therefore be thought of as the intersection of $L\R^n$ with the various  components $T\R^n$ of the n-fold Whitney sum.  This is why the new momentum coordinates $P_i$ never vanish on the subbundle $\B_1$ -  they represent linear frames, and the vectors comprising a linear frame can never vanish identically, and not simply collections of arbitrary  vectors.   Although the example quantizations discussed in section 6 were described in terms of $\B_1$, it is clear that any of the subbundles $\B_i$ could equally well be used with   appropriate modifications of the quantization map.  We were able to define full polynomial quantizations on these 2n-dimensional subbundles because the reduction of the n-symplectic algebra retained an ideal structure, something that is not present in the polynomial algebra defined by symplectic geometry on $T^*\R^n$.

 There are many questions about n-symplectic quantization that still need to be answered. Are there other possible polynomial quantizations that can be defined?  In this paper we have only considered the symmetric set of observables $SHF$.  Are there other possible polynomial quantizations that can be defined using the symmetric observables $SHF$? Can one quantize the antisymmetric observables $AHF$?   Can the n-symplectic analogues of other symplectic manifolds that exhibit obstructions be quantized using n-symplectic quantization?  The authors hope to address these topics in future papers.

\newpage

\section{APPENDIX}

  The following table displays the first few polynomial observables and their corresponding Hamiltonian vector fields.  When appropriate the  n-symplectic gauge freedom has been used   on the vertical component as described in lemma (\ref{TheLemma}) in section 6.   In the table all lower case latin indices run from 1 to $n$.

\begin{table}[htdp]
\caption{Some polynomial observables and their Hamiltonian vector fields}
\begin{center}
\begin{tabular}{|c|c|c|c|} \hline
\# & $\hat f = f^{P_\alpha}\hat r_{\alpha}$& $f^{P_\alpha}$& Hamiltonian vector field(s) \\ \hline
1 & $ \f = \q^i_1 $    & $q^i$&  $ X_{\q^i_1 } = - \basispi 1i   $  \\  \hline
2& $ \f = \q^i_1\r_1 $    & $q^i\delta^a_1\delta^b_1$&  $ X^a_{\q^i_1\r_1 } = - \frac 12 \delta^a_1\basispi 1i    $  \\ \hline
3 & $ \f = \hat\pi_k $    & $\pi_k^j$&  $ X_{\pi_k } =  \basisq k  $  \\ \hline
4 & $ \f = \hat\pi_k \r_1 $    & $\delta_1^{(a}\pi^{b)}_k$&  $ X^a_{\hat\pi_k\r_1 } = \frac 12 \delta^a_1\basisq k  $  \\ \hline
5 & $ \f = \q^i_1 \q^j_1 $    &   $q^iq^j\delta^a_1\delta^b_1   $    &  $  X^a_{\q^i_1 \q^j_1} = -\frac 12 (q^i\basispi 1j + q^j\basispi 1i)\delta^a_1  $  \\ \hline
6 & $\f = \hat\pi_j\p_k  $    &   $ \pi^{(a}_j\pi^{b)}_k $   &  $  X^a_{\hat\pi_j\p_k } =  \frac 12( \pi^a_i\basisq j + \pi^a_j\basisq i ) $  \\ \hline
7 & $ \f = \q^i_1 \hat\pi_k $    &  $ q^i\delta^{(a}_1\pi_k^{b)} $  &  $  X^{a}_{\q^i_1 \hat\pi_k} = \frac 12 q^i\delta^a_1\basisq k -\frac 12 \pi^a_k\basispi 1i   $  \\ \hline
8 & $ \f = \q^i_1 \q^j_1\q^k_1 $    &  $  q^iq^jq^k\delta^a_1\delta^b_1 \delta^c_1 $   &   $ X^{ab}_{\q^i_1 \q^j_1\q^k_1} = -\frac 1{3!} (q^iq^k\basispi 1j + q^jq^k\basispi 1i  + q^iq^j\basispi 1k)\delta^a_1 \delta^b_1  $  \\ \hline
9 & $ \f = \q^i_1 \q^j_1 \hat\pi_k $    &   $q^iq^j\delta^{(a}_1\delta^b_1\pi^{c)}_k    $     &  $  X^{ab}_{ \q^i_1 \q^j_1 \hat\pi_k} = \frac 1{3!} \left( q^iq^j\delta^{(a}_1\delta^{b)}_1\basisq k  - \delta^{(a}_1\pi^{b)}_k \left(q^j\basispi 1i  + q^i\basispi 1j\right) \right)  $  \\ \hline
\end{tabular}
\end{center}
\label{default}
\end{table}%
\bigskip

\newpage

\bibliographystyle{plain}
\bibliography{bibfiles3}

\begin{thebibliography}{10}

\bibitem{JDB}
Jonathan~D. Brown.
\newblock {\em N-Symplectic Quantization}.
\newblock PhD thesis, North Carolina State University, 2008.

\bibitem{DC}
Daniel Cartin.
\newblock Some aspects of geometric pre-quantization on the affine frame
  bundle.
\newblock Master's thesis, North Carolina State University, 1995.

\bibitem{GeomS}
M.~{de Le{\'o}n}, M.~{McLean}, L.~K. {Norris}, A.~{Rey-Roca}, and M.~{Salgado}.
\newblock {Geometric Structures in Field Theory}.
\newblock {\em ArXiv Mathematical Physics e-prints}, aug 2002.

\bibitem{obstructions}
Mark~J. Gotay.
\newblock Obstructions to quantization.
\newblock In {\em Mechanics:From Theory to Computation}. Springer, 2000.

\bibitem{Gr}
H.~J. Groenewold.
\newblock On the principles of elementary quantum mechanics.
\newblock {\em Physics}, 12:405--458, 1946.

\bibitem{STP}
V.~Guillemin and S.~Sternberg.
\newblock {\em Symplectic techniques in physics}.
\newblock Cambridge University Press, 1984.

\bibitem{Gunther}
Ch. Gunther.
\newblock The polysymplectic hamiltonian formalism in field theory and calculus
  of variations i: The local case.
\newblock {\em J. Differential Geometry}, 25:25--53, 1987.

\bibitem{VHove}
Leon~Van Hove.
\newblock {\em On Certain Unitary Representations of an Infinite Group of
  Transformations}.
\newblock World Scientific, 2001.

\bibitem{LKN1}
L.K. Norris.
\newblock Generalized symplectic geometry on the frame bundle of a manifold.
\newblock {\em Symposium in Pure Mathematics}, 54:435--465, 1993.

\bibitem{LKN2}
L.K. Norris.
\newblock {Symplectic Geometry on $T^*M$ Derived from n-symplectic Geometry on
  $LM$}.
\newblock {\em Geometry and Physics}, 13:51--78, 1994.

\bibitem{LKN3}
L.K. Norris.
\newblock {Schouten-Nijenhuis Brackets}.
\newblock {\em Journal of Mathematical Physics}, 38:2694--2709, 1997.

\bibitem{LKN4}
L.K. Norris.
\newblock {The n-symplectic algebra of observables in covariant lagrangian
  field theory}.
\newblock {\em Journal of Mathematical Physics}, 42:4827--4845, 2001.

\end{thebibliography}

\end{document}